\documentclass[amsmath,reprint,aip,apl]{revtex4-1}

\usepackage{graphicx}
\usepackage{hyperref}
\begin{document}

\title
{Cold exciton electroluminescence from air-suspended carbon nanotube split-gate devices}
\author{N.~Higashide}
\affiliation{Department of Electrical Engineering, The University of Tokyo, Tokyo 113-8656, Japan}
\author{M.~Yoshida}
\affiliation{Nanoscale Quantum Photonics Laboratory, RIKEN, Saitama 351-0198, Japan}
\affiliation{Quantum Optoelectronics Research Team, RIKEN Center for Advanced Photonics, Saitama 351-0198, Japan}
\author{T.~Uda}
\affiliation{Nanoscale Quantum Photonics Laboratory, RIKEN, Saitama 351-0198, Japan}
\affiliation{Department of Applied Physics, The University of Tokyo, Tokyo 113-8656, Japan}
\author{A.~Ishii}
\affiliation{Department of Electrical Engineering, The University of Tokyo, Tokyo 113-8656, Japan}
\affiliation{Nanoscale Quantum Photonics Laboratory, RIKEN, Saitama 351-0198, Japan}
\author{Y.~K.~Kato}
\email[Author to whom correspence should be addressed. Electronic mail: ]{yuichiro.kato@riken.jp}
\affiliation{Nanoscale Quantum Photonics Laboratory, RIKEN, Saitama 351-0198, Japan}
\affiliation{Quantum Optoelectronics Research Team, RIKEN Center for Advanced Photonics, Saitama 351-0198, Japan}

\begin{abstract}
Electroluminescence from individual carbon nanotubes within split-gate devices is investigated. By characterizing the air-suspended nanotubes with photoluminescence spectroscopy, chirality is identified and electroluminescence peaks are assigned. We observe electroluminescence linewidth comparable to photoluminescence, indicating negligible heating and state-mixing effects. Split-gate and bias voltage dependences are consistent with emission from an electrostatically formed $pn$-junction.
\end{abstract}

\maketitle

Single-walled carbon nanotubes (CNTs) offer unique characteristics for applications in nanoscale optoelectronics, being a one-dimensional semiconductor that can be directly grown on silicon. With their diameters of a few nm or less, they can be utilized as near field emitters, while their lengths can be over many microns, allowing for straightforward fabrication of electrically driven devices. Consequently, electroluminescence (EL) from CNTs have been investigated in various device structures,\cite{Freitag:2004prl, Chen:2005, Mann:2007, Mueller:2010, Pfeiffer:2011, Wang:2011, Xie:2012acs, Jakubka:2014} and numerous mechanisms have been proposed to interpret the experimental results. Nevertheless, typical EL spectra exhibit broad linewidths of over 100~meV,\cite{Freitag:2004nl, Chen:2005, Mann:2007, Pfeiffer:2011, Xie:2012acs} indicative of significant heating caused by bias voltages required to drive the devices. Such a spectral broadening is undesirable for coupling to high quality-factor cavities\cite{Watahiki:2012, Imamura:2013, Miura:2014, Noury:2015, Pyatkov:2016, Jeantet:2016} and it is also known that quantum efficiencies decrease at higher temperatures.\cite{Mortimer:2007} In this context, a promising result has been reported for electrostatically formed $pn$-junctions using split-gate devices, where a linewidth as narrow as 35~meV was observed from a nanotube embedded between a substrate and a gate dielectric.\cite{Mueller:2010}  Further investigation of split-gate devices, in particular those utilizing air-suspended tubes, have revealed ideal diode behavior and superior photoconductivity performance.\cite{Lee:2005, Lee:2007prb, Liu:2011nl, Amer:2013, Barkelid:2014, DeBorde:2014} Considering these experimental efforts on electrostatically formed $pn$-junctions in air-suspended CNTs, improved EL performance is expected as well.

Here we report on EL from air-suspended carbon nanotube split-gate devices, and show that EL can be obtained without any spectral broadening in comparison to photoluminescence (PL). By characterizing the nanotubes with PL excitation spectroscopy, nanotube chirality is identified and EL is assigned to $E_{11}$ exciton recombination. Split-gate and bias voltage dependences of EL intensity are both consistent with emission originating from an electrostatically formed $pn$-junction, and the reconfigurable nature of split-gate devices is also demonstrated.

\begin{figure}
\includegraphics{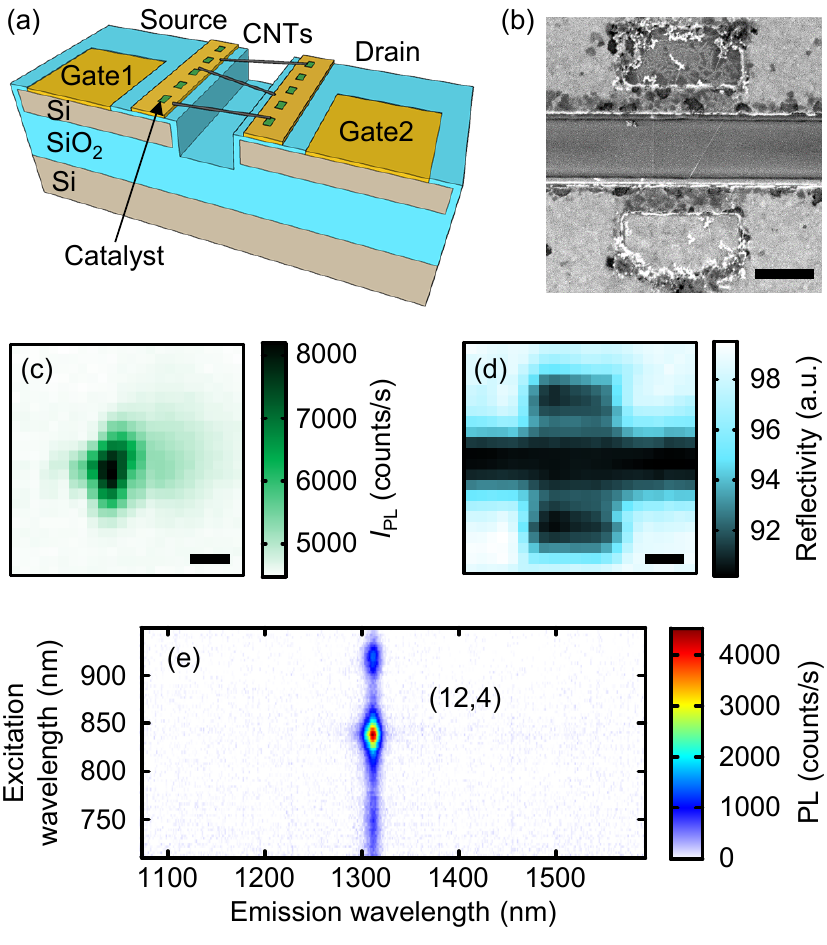}
\caption{\label{fig1}
(a)~Schematic of a carbon nanotube field-effect split-gate device. (b)~Scanning electron micrograph of a typical device. (c) and (d)~PL and reflectivity images, respectively, in a device with a nanotube showing EL. Data are taken with an excitation power $P=30~\mu$W and an excitation wavelength of 836~nm. For (c), the spectral integration window for the PL image is from 1304 to 1330~nm. (e)~PL excitation map of the nanotube measured in (c), and $P=50~\mu$W is used. In (b-d), the scale bars are 1~$\mu$m. For (c-e), the laser polarization is perpendicular to the trench.
}\end{figure}

A schematic of our split-gate field-effect device is shown in Fig.~\ref{fig1}(a). Air-suspended CNTs are contacted on both sides of a trench, and two local gates are used for electrostatic doping. The devices are fabricated on silicon-on-insulator substrates with a 260-nm-thick top Si layer and 1-$\mu$m-thick buried oxide.\cite{Jiang:2015} The top Si layer is boron doped with a resistivity of 18.0$\pm$4.5~$\Omega$~cm, allowing for use as split gates. We begin by performing electron beam lithography to define trenches with a width of 1~$\mu$m for suspending the nanotubes, and etch 1000~nm deep through the top Si layer and into the buried oxide with an inductively coupled plasma etcher with CHF$_3$ gas. Thermal oxidation is then performed at 1050$^\circ$C for 30~minutes to form a 55-nm-thick SiO$_2$ layer on the top Si layer, and another electron beam lithography step patterns the split-gate electrodes. A wet etching process removes $\sim$40~nm of thermal oxide, and we deposit Ti (2~nm)/Pt (20~nm) by electron beam evaporation for contacting the top Si layer. Following a lift-off process of the gate electrodes, source and drain contacts are patterned right next to the trenches by a third electron beam lithography step. We evaporate SiO$_2$ (80~nm)/Ti (2~nm)/Pt (20~nm) and perform lift-off to form the contacts. A fourth lithography step defines the catalyst areas overlapping small windows within the source and drain contacts, and Fe/silica dissolved in ethanol is spin-coated and lifted-off.\cite{Ishii:2015} Finally, CNTs are synthesized by ethanol chemical vapor deposition at 750$^\circ$C for 1~minute. Figure~\ref{fig1}(b) shows a scanning electron micrograph of a typical device.

The devices are characterized with a home-built microspectroscopy system, where an automated three-dimensional stage is utilized for sample scanning.\cite{Uda:2016, Jiang:2015, Ishii:2015} For EL measurements, the source contact and the substrate are grounded, and we apply voltages $V_{\text{G1}}$ to gate~1, $V_{\text{G2}}$ to gate~2, and $V_{\text{DS}}$ to the drain contact. As an excitation source for PL measurements, a wavelength-tunable continuous wave Ti:sapphire laser is used, and an objective lens with a numerical aperture of 0.8 and a focal length of 1.8~mm focuses the laser on the sample. Light emission from the sample is collected with the same objective lens, and a 50-mm focal length lens is used to couple to a 300-mm spectrometer which houses a 150-lines/mm grating for spectroscopy and a mirror for imaging. The emission signals are detected by a 25.6-mm-long 512-pixel InGaAs photodiode linear array. Background subtraction is performed with data taken at all gate and bias voltages at zero for EL measurements, while a laser beam shutter is utilized for PL spectroscopy. Samples are kept in N$_2$ atmosphere to reduce gate hysteresis,\cite{Kim:2003} and all measurements are performed at room temperature. 

In order to find suspended nanotubes that show EL, we perform one-dimensional imaging measurements by using the mirror in the spectrometer. Linear images with a field-of-view of $\sim$50~$\mu$m are acquired at voltages of $V_{\text{G1}}=V$, $V_{\text{G2}}=-V$, and $V_{\text{DS}}=V/2$, and each 610-$\mu$m-long trench is scanned by collecting such images in 50-$\mu$m steps. If no emission is observed, we repeat the scan with an increased $V$ until we find a signal or $V$ reaches 30~V. 

Once a light-emitting nanotube is identified, we turn off the voltage and characterize the tube by PL spectroscopy. A confocal pinhole with a diameter of 150~$\mu$m is placed at the entrance of the spectrometer, and the mirror in the spectrometer is switched to the grating. In Fig.~\ref{fig1}(c), a PL image obtained by mapping out the integrated PL intensity $I_\text{PL}$ is shown. By comparing to a reflectivity image [Fig.~\ref{fig1}(d)], we confirm that the nanotube is fully suspended over the trench. In order to identify the chirality, PL excitation spectroscopy is performed with a laser power $P=50~\mu$W [Fig.~\ref{fig1}(e)], and we determine the tube chirality to be (12,4) using tabulated data.\cite{Ishii:2015} The laser polarization dependence of PL is also measured to determine the angle of the tube.\cite{Moritsubo:2010} These measurements ensure that EL spectroscopy is performed on a semiconducting nanotube that is fully suspended and individual. We note that some nanotubes do not show PL, but exhibit electrically induced emission with broad linewidths on the order of 100~meV. We attribute such emission to thermal light emission from metallic CNTs.\cite{Mann:2007}

\begin{figure}
\includegraphics{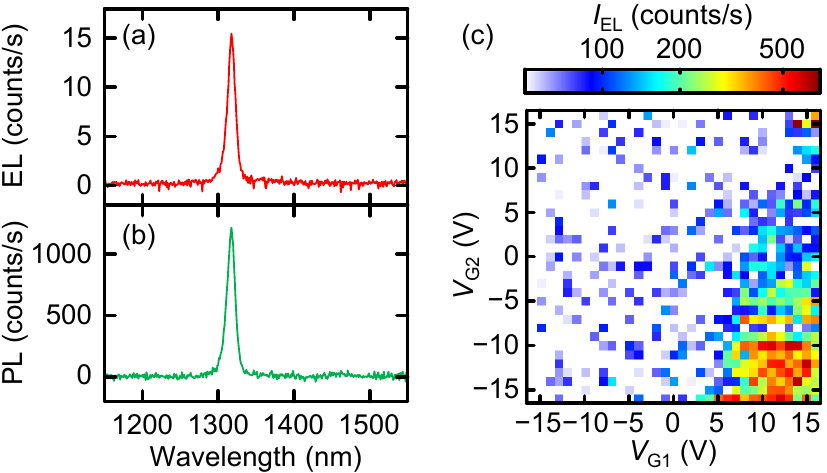}
\caption{\label{fig2}
Light emission from the nanotube characterized in Figs.~\ref{fig1}(c-e).
(a)~EL spectrum taken at $V_\text{G1}=16$~V, $V_\text{G2}=-16$~V, and $V_\text{DS}=5$~V. (b)~PL spectrum taken with $V_\text{G1}=V_\text{G2}=V_\text{DS}=0$~V and $P=5~\mu$W. The laser polarization
angle is adjusted to maximize the PL signal, and a resonant excitation wavelength of 836~nm is used. (c)~Integrated EL as a function of $V_\text{G1}$ and $V_\text{G2}$ taken with $V_\text{DS}=7$~V.
}\end{figure}

On the (12,4) tube characterized in Figs.~\ref{fig1}(c-e), we have obtained an EL spectrum under an application of voltages $V_\text{G1}=16$~V, $V_\text{G2}=-16$~V, and $V_\text{DS}=5$~V [Fig.~\ref{fig2}(a)]. The peak emission wavelength and the linewidth are identical to those for a PL spectrum taken in the absence of the voltages [Fig.~\ref{fig2}(b)], showing that the EL peak originates from the $E_{11}$ excitons. The observed full-width at half-maximum of the emission is about 8~meV, which is significantly smaller than previous reports on EL linewidths ranging from 35~meV  to $\sim$100~meV.\cite{Freitag:2004nl, Chen:2005, Mueller:2010, Pfeiffer:2011, Wang:2011, Xie:2012acs, Jakubka:2014} Broadening of emission has been attributed to high carrier temperatures as well as field-induced mixing of excitonic states with the continuum,\cite{Mueller:2010} and in particular heating-induced line broadening has been well characterized by PL spectroscopy.\cite{Yoshikawa:2009} The sharp peak observed in Fig.~\ref{fig2}(a) indicates that excitons remain cold, with negligible Joule heating effects keeping the nanotube in thermal equilibrium with the substrate. In addition, the data also demonstrates that EL can be obtained at modest bias and internal fields where state mixing effects do not play a role.

We note that EL intensity [Fig.~\ref{fig2}(a)] is considerably smaller compared to PL intensity [Fig.~\ref{fig2}(b)], which may be caused by differences in the length of luminescent regions. Emission intensity quenches with an application of gate voltage,\cite{Yasukochi:2011, Jiang:2015, Yoshida:2016} and only a limited length would remain active in a $pn$-junction configuration. In comparison, the entire suspended length of the nanotube is luminescent for PL measurements. Another likely reason is the difference in the excitation powers for EL and PL. Unfortunately, we are not able to determine the current through this nanotube as our device structure has multiple tubes, making it difficult to quantify the electrical excitation power and the external quantum efficiency. In addition, there exists uncertainties for absorption cross section of nanotubes,\cite{Joh:2011, Oudjedi:2013, Malapanis:2013, Streit:2014, Kumamoto:2014, Ishii:2015} causing complications in quantifying optical excitation rates as well.  

In order to confirm the role of electrostatically formed $pn$-junction, we investigate EL dependence on $V_\text{G1}$ and $V_\text{G2}$ at a constant bias $V_\text{DS}=7$~V [Fig.~\ref{fig2}(c)], where spectrally integrated EL intensity $I_\text{EL}$ is collected by switching the grating to a mirror. We observe bright emission only in the bottom-right corner of the map, corresponding to the region with $V_\text{G1}>0$~V and $V_\text{G2}<0$~V where a forward biased $pn$-junction should be formed. In the top-left corner of the map, a $pn$-junction should also exist but it would be reverse biased, and in the top-right and the bottom-left regions of the map, there should not be any $pn$-junction as $V_\text{G1}$ and $V_\text{G2}$ have the same polarity. The experimental results are mostly consistent with such a picture, and slight deviations may arise from residual hysteretic effects of the gates. We note that EL linewidth did not show any detectable change for different gate voltage combinations.

\begin{figure}
\includegraphics{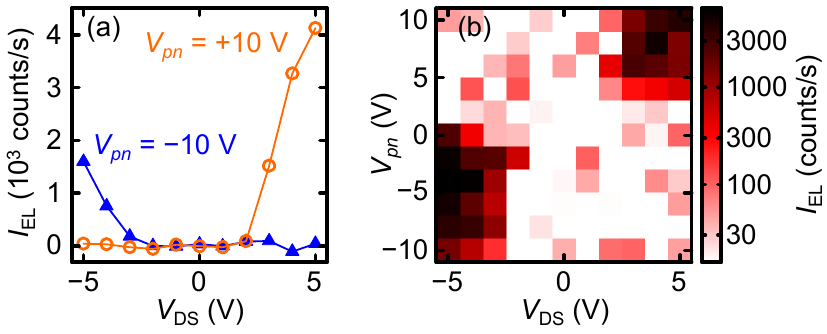}
\caption{\label{fig3}
(a)~Drain voltage dependence of integrated EL intensity for a (9,8) nanotube with $V_{pn}=10$~V (orange open circles) and $-10$~V (blue filled triangles). (b)~Integrated EL as a function of $V_\text{DS}$ and $V_{pn}$.
}\end{figure}

Next, we examine the $V_\text{DS}$ dependence for another tube, where we define $V_{pn}=V_\text{G1}=-V_\text{G2}$ to be the symmetrically-applied gate voltage that forms the $pn$-junction. The bias dependence of EL [Fig.~\ref{fig3}(a)] resembles a current-voltage curve for a $pn$-junction diode, which is reasonable as emission should be proportional to the current. EL is observed for both polarities of $V_\text{DS}$ for this device, if the sign of $V_{pn}$ is also inverted to form a forward biased $pn$-junction. We note that not all tubes show EL for both polarity of $V_\text{DS}$, presumably due to asymmetries in the contact resistance. A detailed map of gate and bias dependence [Fig.~\ref{fig3}(b)] shows that this device can be switched from a $pn$-junction to a $np$-junction configuration with a change of few volts, demonstrating the reconfigurability of the split-gate devices. 

In summary, we have investigated EL from suspended carbon nanotubes within split-gate devices. By characterizing the same nanotube by PL spectroscopy, emission has been attributed to $E_{11}$ exciton recombination. We observe EL emission linewidth comparable to those for PL, indicating that EL can be obtained without heating up the nanotube and with negligible field-induced state mixing. Split-gate voltage and bias voltage dependences show that emission occurs for voltage combinations corresponding to forward-biased $pn$-junctions, confirming the flexible configuration of the device utilizing electrostatic carrier doping.

\begin{acknowledgments}
Work supported by JSPS (KAKENHI JP16H05962, JP26610080), MEXT (Photon Frontier Network Program, Nanotechnology Platform), Canon Foundation, and Asahi Glass Foundation. T.U. is supported by ALPS, and A.I. is supported by MERIT and JSPS Research Fellowship.
\end{acknowledgments}

\end{document}